\documentclass[conference]{IEEEtran}
\IEEEoverridecommandlockouts
\usepackage{cite}
\usepackage{array}
\usepackage{tabularx}
\usepackage{amsmath,amssymb,amsfonts}
\usepackage{algorithmic}
\usepackage{graphicx}
\usepackage{textcomp}
\usepackage{xcolor}
\usepackage[disable]{todonotes}
\setuptodonotes{inline}
\usepackage{hyperref}
\usepackage{cleveref}
\usepackage{booktabs}
\def\BibTeX{{\rm B\kern-.05em{\sc i\kern-.025em b}\kern-.08em
    T\kern-.1667em\lower.7ex\hbox{E}\kern-.125emX}}

\makeatletter
\g@addto@macro{\UrlBreaks}{\UrlOrds} %
\makeatother

\begin{document}

\title{
Neuromorphic hardware for sustainable AI data centers
\thanks{This work was partially funded by the German Federal Ministry for Economic Affairs and Climate Action (BMWK) under contracts
01MN23004A, %
01MN23004D, %
01MN23004F, %
and 01MN23994B %
and by the German Federal Ministry of Education and Research (BMBF) and the free state of Saxony within the ScaDS.AI center of excellence for AI research.
This work was partially funded by the EIC Transition program under the “SpiNNode” project (grant number 101112987). %
}
}

\author{\IEEEauthorblockN{
Bernhard Vogginger\IEEEauthorrefmark{1}\IEEEauthorrefmark{11},
Amirhossein	Rostami\IEEEauthorrefmark{1}\IEEEauthorrefmark{6},
Vaibhav	Jain\IEEEauthorrefmark{4},
Sirine Arfa\IEEEauthorrefmark{1},
Andreas	Hantsch\IEEEauthorrefmark{7}\IEEEauthorrefmark{8},
David Kappel\IEEEauthorrefmark{5},
\\
Michael Schäfer\IEEEauthorrefmark{2}\IEEEauthorrefmark{3},
Ulrike Faltings\IEEEauthorrefmark{2},
Hector A. Gonzalez\IEEEauthorrefmark{1}\IEEEauthorrefmark{6}\IEEEauthorrefmark{9},
Chen Liu\IEEEauthorrefmark{1},
Christian Mayr\IEEEauthorrefmark{1}\IEEEauthorrefmark{6},
Wolfgang Maaß\IEEEauthorrefmark{4}\IEEEauthorrefmark{10}
}
\IEEEauthorblockA{\IEEEauthorrefmark{1}Chair of Highly-Parallel VLSI-Systems and Neuro-Microelectronics, Technische Universität Dresden, Germany}
\IEEEauthorblockA{\IEEEauthorrefmark{2}SHS - Stahl-Holding-Saar GmbH \& Co. KGaA, Germany}
\IEEEauthorblockA{\IEEEauthorrefmark{3}KTH Royal Institute of Technology, Department of Materials Science and Engineering, Sweden}
\IEEEauthorblockA{\IEEEauthorrefmark{4}German Research Center for Artificial Intelligence (DFKI), Saarbrücken, Germany}
\IEEEauthorblockA{\IEEEauthorrefmark{5}Institute for Neural Computation, Ruhr University Bochum, Germany}
\IEEEauthorblockA{\IEEEauthorrefmark{6}ScaDS.AI Dresden/Leipzig, Germany}
\IEEEauthorblockA{\IEEEauthorrefmark{7}eco2050 Institut für Nachhaltigkeit – Institute for Sustainability GmbH, Nürnberg, Germany}
\IEEEauthorblockA{\IEEEauthorrefmark{8}Hantsch Sustainability Consulting, Dresden, Germany}
\IEEEauthorblockA{\IEEEauthorrefmark{9}SpiNNcloud Systems GmbH, Dresden, Germany}
\IEEEauthorblockA{\IEEEauthorrefmark{10}Saarland University, Saarbrücken, Germany}
\IEEEauthorblockA{\IEEEauthorrefmark{11}Email: bernhard.vogginger@tu-dresden.de}

}

\maketitle
\thispagestyle{plain}
\pagestyle{plain}

\begin{abstract}
As humans advance toward a higher level of artificial intelligence, it is always at the cost of escalating computational resource consumption, which requires developing novel solutions to meet the exponential growth of AI computing demand.
Neuromorphic hardware takes inspiration from how the brain processes information and promises energy-efficient computing of AI workloads. Despite its potential, neuromorphic hardware has not found its way into commercial AI data centers. In this article, we try to analyze the underlying reasons for this and derive requirements and guidelines to promote neuromorphic systems for efficient and sustainable cloud computing: We first review currently available neuromorphic hardware systems and collect examples where neuromorphic solutions excel conventional AI processing on CPUs and GPUs. Next, we identify applications, models and algorithms which are commonly deployed in AI data centers as further directions for neuromorphic algorithms research. Last, we derive requirements and best practices for the hardware and software integration of neuromorphic systems into data centers. With this article, we hope to increase awareness of the challenges of integrating neuromorphic hardware into data centers and to guide the community to enable sustainable and energy-efficient AI at scale.
\end{abstract}


\section{Introduction}
\subsection{Motivation}
\todo{R1: One of my main concerns with the work is that it is challenging to make a fair comparison across the hardware implementations because of the lack of common benchmarks that are included (which is something they note in the paper)}
\todo{R1: input output coding is a much bigger challenge and will likely consume quite a bit of computation}
\todo{R2: I have concerns that the field is moving too quickly for this paper to be relevant for long}
\todo{R2: In addition, the paper doesn’t add its own independent benchmarking, it simply summarizes the publicly available benchmarking for these hardware systems and calls for additional benchmarking to be done}
\todo{R2: Finally, most of the recommendations for the field presented in the paper seem obvious and I don't feel like the paper spends enough time digging into the challenges mentioned and suggesting potential solutions}
\todo{R3: additional supporting work is needed for Table 2 etc. --> add appendix}
\todo{R3: No discussion of what is sustainable in this context}
\todo{R3: What are the costs to replacing existing infastructure with neuromorphic systems - how much of general purpose computing can be replaced with neuromorphic because they offer more than just AI - what are the expected impacts}
\todo{R3: Mention of few examples of where neuromorphic systems have been integrated into data centers - what were the results/performance}
\todo{R3: Grammar should be checked}
\todo{R3: The challenges are presented at the end but there is no discussion around these challenges - what is the severity of them (where we are now -> where we need to be), their impact and ranking in preventing integration into data centers, if these are the key challenges and why.}
\todo{Fabrizio-1: provide Accuracy, energy per inference, throughput, model size, timestep -> try to add model size and timesteps to the appendix}
\todo{Fabrizio-2: - An Loihi2 (as Loihi 1) is not commercially available
- And in the comparison between Loihi2 and Jetson, it seems that you considered the data without IO (which is DRAM), because Jetson would be much faster otherwise (also expected)
- And the Jetson has a better EDP}
\todo{Fabrizio-3: But comparing SNN accelerators to 400W GPUs is not super fair}
\todo{Fabrizio-4: Anyway, one should compare with ANN inference accelerators instead of GPUs, in my opinion}
\todo{alex: Neuromorphic Circuits will be incorporated in SoCs oder MSoCs in the Future. Could be a nice discussion in this chapter. 
What is the Status quo (check)
What would be a viable way in the future}

Data centers, serving as the hub for computers and equipment necessary to manage and store vast amounts of data, play a crucial role in deploying and maintaining AI systems, especially given the exponential growth in demand for AI computing models \cite{wu2022sustainable}.
However, advances in traditional computers (CPU, GPU, TPU) have not kept pace with this growing demand. There is thus an urgent need for innovative solutions across hardware, software, and algorithms to ensure efficient, high-throughput, and sustainable AI in data centers~\cite{guitart2017toward}. 

Among potential candidates, neuromorphic computing inspired by the human brain is emerging as a promising approach to address these challenges with the feature of energy-efficient parallel processing \cite{roy2019towards,rathi2023exploring}.
Neuromorphic computing aims to design and build computer systems including hardware and software that can perform cognitive tasks more efficiently by emulating how neurons and synapses work in the brain.
Neuromorphic systems incorporate the concept of synaptic plasticity, allowing synapses between spiking neurons to change and adapt based on the input patterns. 
This work addresses the integration of neuromorphic hardware into data centers for sustainable AI. 

\subsection{Sustainable AI}
Data centers have a tremendous energy demand. Whilst the estimates vary over the order of a magnitude, their median global electricity demand was 300 TWh/a in 2020 and almost tripled by 2030~\cite{mytton_2022_Joule_6_2032}. %

For a long period, just the energy demand as part of the environmental impact was in focus~\cite{wu2022sustainable, hantsch_2021_ocp}. However, the term \textit{"sustainable AI"} describes the creation and application of AI technologies that prioritize long-term viability, social responsibility, and reducing environmental impact. As these applications spread across various industries, there's a rising awareness of the necessity to address ethical and environmental issues related to AI development and implementation.

Following the Environment, Social, and Governance (ESG) structure of financial ratings, we highlight the following impact points for sustainable AI:

\begin{itemize} %
    \item \textbf{Environment:} 1) Materials, water, land, refrigerants, and energy (through greenhouse gases) needed during the server and data center life cycles - including embodied carbon dioxide; 2) Computing demand for AI training and inference computations
    \item \textbf{Social:}  1) Gathering (and labeling) of unbiased, racist-free training data embracing human rights; 2) Transparency and explainability of algorithms; 3) Potential positive impact on society by well-designed AI applications
    \item \textbf{Governance:} 1) Data security; 2) Data and service availability with shifting context, data sets, and model features; 3) Virtualisation and load balancing for operation optimization; 4) Economic feasibility
\end{itemize}
We would like to highlight that the embodied carbon dioxide in IT hardware can be more than half of the total carbon footprint of this hardware's life cycle~\cite{mcgovern_2020_www}. 
Yet, in this article, we focus on reducing the energy demand for AI processing in data centers by deploying neuromorphic computing hardware and algorithms.

\subsection{Related work}

This work addresses a closely related issue to the task explored in \cite{osti_1473756}, which involves integrating diverse emerging hardware systems, including neuromorphic systems (NC), into a unified computational environment. The authors of \cite{nilsson2023integration} emphasize the significance of computational environments, covering conventional digital computing (DC) systems based on the von Neumann architecture and synchronous logical processing, alongside traditional distributed computing. They define NC as event-based systems with a distinct interface, wherein the structure and function either emulate or simulate the neuronal dynamics of brains, particularly somas, and occasionally synapses, dendrites, and axons, typically represented in the form of Spiking Neural Networks (SNNs). The study highlights the interconnection between the two types of hardware, DC and NC, establishing a microservice-based conceptual framework for integrating neuromorphic systems, featuring a neuromorphic-system proxy.
\newline
In contrast to the approach in \cite{nilsson2023integration}, %
we are currently placing a greater emphasis on promoting the adoption of neuromorphic systems for typical AI tasks within data centers, with the aim of optimizing efficiency in cloud computing. This shift in focus directs our attention away from the generation and decoding aspects and principles in hardware associated with event processing.

Recently, significant progress has been made in neuromorphic computing, particularly in the realm of SNNs. In \cite{rathi2023exploring} the authors highlighted the challenges that must be overcome within this field to fully leverage the potential of efficient AI computing. The demands placed on SNN accelerators have witnessed notable transformations, especially in the design of large-scale systems capable of effectively leveraging the essential features of SNN algorithms. The authors emphasized the design principles of neuromorphic hardware architectures, drawing inspiration from two core tenets of SNNs: (i) event-driven sparse computations and (ii) efficient and parallel matrix operations.
A comparison between these neuromorphic hardware architectures and the standard architectures for Artificial Neural Networks (ANNs) revealed the proficiency of the standard architectures in matrix operations but their shortfall in exploiting the temporal sparsity inherent in SNNs.

Similar to this work, \cite{luo2023achieving} discusses the potential of neuromorphic hardware for energy-efficient and green AI computing. The approach is showcased with the Novena chip from Singapore. The complete eco-system for deploying the chip is discussed, including algorithms, software, middleware, and system integration. Instead, our work analyses the current neuromorphic computing landscape as well as the potential and challenges for the integration of NC into mainstream AI data centers. 

\subsection{Outline}

This paper considers key participants in the neuromorphic systems field, including both commercial entities and major academic contributors, in the context of AI data centers and broader solutions. We present a detailed analysis of these systems and highlight the importance of hardware and software integration. The paper showcases the need for a holistic perspective by comparing neuromorphic to conventional approaches for AI processing. 

Furthermore, it identifies common AI tasks in data centers and serves as a valuable reference for researchers investigating the development of neuromorphic algorithms. Fundamentally, the overall goal of this initiative is to improve the sustainability of data centers.

\section{Neuromorphic Hardware Players}
\label{sec:neuromorphic}

\begin{table*}
\centering
\renewcommand{\arraystretch}{1.3}
\caption{Overview of established neuromorphic systems from industry and academia}
\begin{tabular}{lllllllll} 
\toprule
\textbf{System}        & \textbf{Developer}                                                               & \textbf{Technology} & \textbf{System size} & \textbf{Neurons/chip} & \textbf{Synapses/chip} & \textbf{Training} & \textbf{Has CPU} & \textbf{Framework}                                                      \\ 
\hline
\textbf{SNP T1}        & Innatera                                                                         & Mixed (28nm)        & Edge                 & 1K                    & -                      & $\times$                 & \checkmark            & Talamo                                                                  \\
\textbf{Speck}         & SynSense                                                                         & Digital (65nm)      & Edge                 & 320K                  & -                      & $\times$                 & $\times$                & Sinabs                                                                  \\
\textbf{Xylo}          & SynSense                                                                         & Digital (40nm)      & Edge                 & 1K                    & 64K                    & $\times$                 & $\times$                & Rockpool                                                                \\
\textbf{DYNAP-SE2}     & SynSense                                                                         & Mixed (180nm)       & Edge                 & 1K                    & 65K                    & \checkmark                & $\times$                & Rockpool                                                                \\
\textbf{Akida}         & BrainChip                                                                        & Digital (28nm)      & Edge/Cloud           & -                     & -                      & \checkmark                & \checkmark               & metaTF                                                                  \\
\textbf{GrAI VIP}      & \begin{tabular}[c]{@{}l@{}}GrAI Matter \\Labs\end{tabular}                       & Digital (28nm)      & Edge                 & 200K                  & -                      & \checkmark                & \checkmark               & GrAIFlow                                                                \\
\textbf{Loihi 1}       & Intel Labs                                                                       & Digital (14nm)      & Edge/Cloud           & 128K                  & 128M                   & \checkmark                & \checkmark               & Lava/NxSDK                                                              \\
\textbf{Loihi 2}       & Intel Labs                                                                       & Digital (Intel 4)   & Edge/Cloud           & 1M                    & 120M                   & \checkmark                & $\times$                & Lava                                                                    \\
\textbf{Tianjic}       & \begin{tabular}[c]{@{}l@{}}Tsinghua \\University\end{tabular}                    & Digital (28nm)      & Edge/Cloud           & 40K                   & 10M                    & $\times$                 & $\times$                & TJSim                                                                   \\
\textbf{BrainScaleS-1} & \begin{tabular}[c]{@{}l@{}}U Heidelberg\end{tabular}                             & Mixed (180nm)      & Cloud                & 197K                   & 43M                    & \checkmark                & $\times$                & PyNN                                                           \\
\textbf{BrainScaleS-2} & \begin{tabular}[c]{@{}l@{}}U Heidelberg\end{tabular}                             & Mixed (65nm)        & Edge/Cloud           & 2K                    & 131K                   & \checkmark                & $\times$                & PyNN, hxTorch                                                           \\
\textbf{NorthPole}     & IBM                                                                              & Digital (12nm)      & Cloud                & 1M                    & 256M                   & $\times$                 & $\times$                & \begin{tabular}[c]{@{}l@{}}NorthPole software \\toolchain\end{tabular}  \\
\textbf{TrueNorth}     & IBM                                                                              & Digital (28nm)      & Edge/Cloud           & 1M                    & 256M                   & $\times$                 & $\times$                & CoreLet (Matlab)                                                        \\
\textbf{SpiNNaker}     & \begin{tabular}[c]{@{}l@{}}U Manchester\end{tabular}                             & Digital (130nm)     & Edge/Cloud           & 16K                   & 16M                    & \checkmark                & \checkmark               & PyNN                                                                    \\
\textbf{SpiNNaker2}    & \begin{tabular}[c]{@{}l@{}}TU Dresden, \\U Manchester\end{tabular}               & Digital (22nm)      & Edge/Cloud           & 152K                  & 152M                   & \checkmark                & \checkmark               & Py-spinnaker2                                                        \\
\bottomrule
\end{tabular}
\label{tab:systems}
\end{table*}

\begin{figure}[htbp]
    \centering
    \includegraphics[width=1\linewidth]{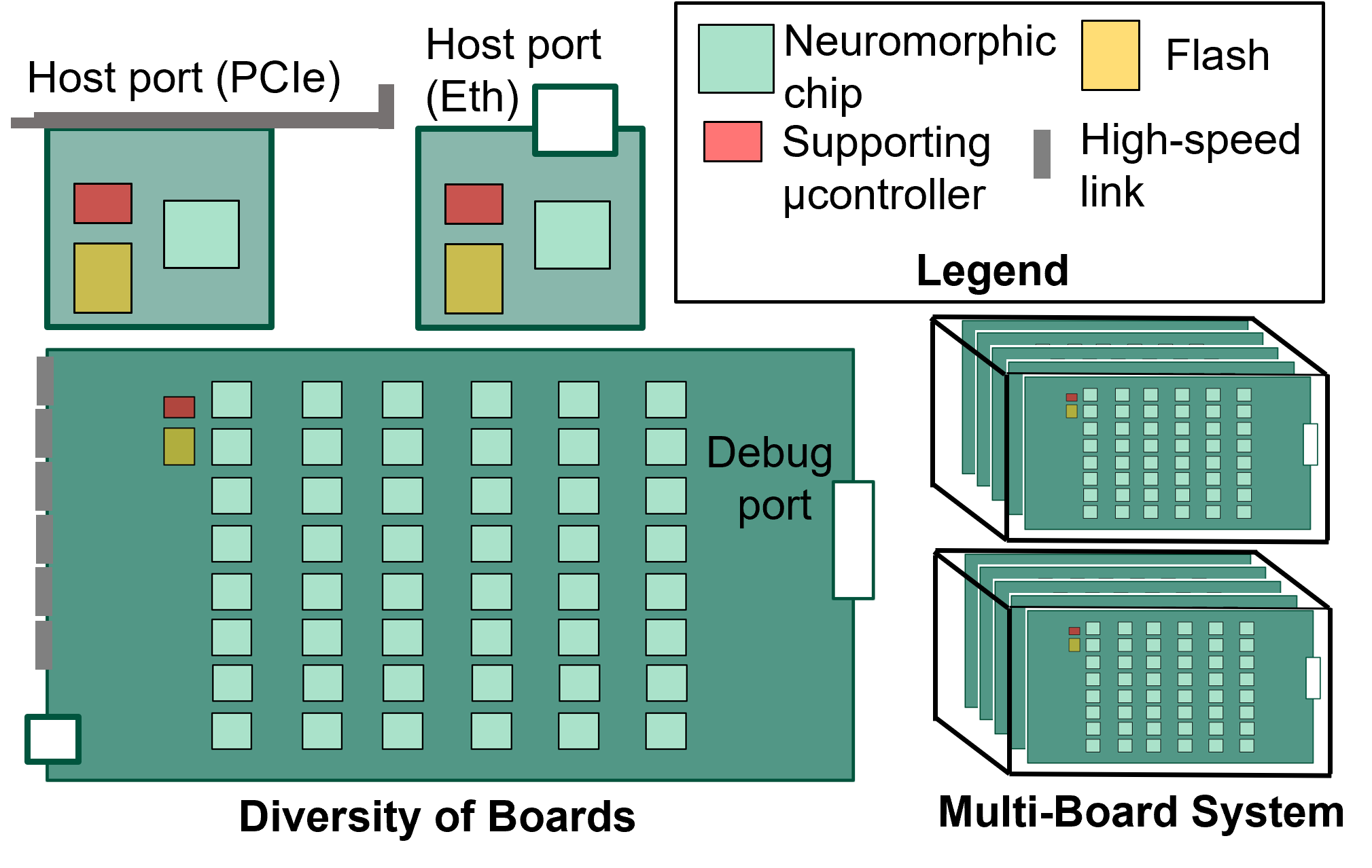}
    \caption{Different scales of neuromorphic systems
    }
    \label{fig:system_types}
\end{figure}

The neuromorphic field currently has a wide diversity of hardware platforms implemented and deployed at different scales aiming to cover applications that range from low-power embedded sensing intelligence \cite{synsensespeck,ward2021innatera} to green cloud services \cite{davies2018loihi,Mayr2019}. Such diversity can be observed via the different board levels in Fig. \ref{fig:system_types}, which presents neuromorphic chips assembled in application boards with PCIe or Ethernet, and in server boards including high-speed links to assemble complex infrastructures to maintain real-time requirements. This section provides a broad overview of some of those neuromorphic platforms with special emphasis on those reaching data center levels.

\subsection{Methodology}
As the term ``neuromorphic'' is used for various kinds of hardware methodologies, we briefly provide a taxonomy of ``neuromorphic systems'' in this paper. We consider \emph{spike-based and event-based} hardware systems in all kinds of implementation approaches (analog, mixed-signal, digital, or processor-based). Most often, those are multi-core architectures where each core combines synapse and neuron processing with a colocation of memory and computation in contrast to von Neumann architectures.
We focus on neuromorphic ASICs and multi-chip systems from both industry and academia with high accessibility and mature software support. We think that these criteria are prerequisites for a near-term integration of such systems into data centers. Neuromorphic architectures such as SENECA \cite{tang2023seneca} or FPGA-based systems like DeepSouth \cite{wang2018fpga} are currently not covered but should be evaluated in the future. 

We neither include neuromorphic compute-in-memory (CIM) architectures that perform vector-matrix multiplication in SRAM, analog crossbars, memristive or other nonvolatile memories such as \cite{wan2022compute,legallo2023AIMC} nor neuromorphic photonics \cite{shastri2021photonics}. While those technologies promise very energy-efficient computing of AI models, they may face other challenges than spiking neuromorphic processors and are thus not part of this review. We also do not consider specialized digital DNN accelerator systems that aim to compete with the GPUs and TPUs. Such DNN accelerator systems are typically much closer integrated into powerful multiprocessor CPUs. But of course, neuromorphic chips should always be compared to the state-of-the-art, which includes digital DNN accelerators.

We also note that our list of hardware platforms is non-exhaustive. We refer to \cite{furber2016large}, \cite{thakur2018large}, and \cite{ivanov2022neuromorphic} for overviews of large-scale neuromorphic systems and projects, and to \cite{basu2022spiking} for a recent overview of trends in SNN processors.
\cite{ottati2023tospike} compares digital SNN and DNN accelerators regarding efficiency and accuracy for DNN workloads.

\subsection{Neuromorphic Systems}
We introduce mature neuromorphic systems sorted by the underlying companies or research groups. A comparison table is available at \cref{tab:systems}.

\subsubsection{Innatera}
The Innatera chip is an analog mixed-signal spiking neural processor for low-power edge applications.
The chip comprises a low-power CPU, encoders, LIF neurons, and programmable synapses.
There are around 1000 analog spiking neurons in each chip, connected through a multi-level crossbar structure \cite{ward2021innatera}.
Due to the limited number of neurons and constrained CPU power, the chip is particularly suitable for processing one-dimensional signals, such as those emerging in audio and healthcare applications.

\subsubsection{SynSense}
There are several commercial neuromorphic chips released by SynSense, such as Dynap-CNN, Speck\cite{yao_spike-based_2024}, Xylo \cite{Bos2022SubmWNS}, and DYNAP-SE2 \cite{richter2024dynapse2}.
The Speck and Dynap-CNN are digital chips designed for real-time vision processing applications such as gesture control, fall detection, and object tracking.
The Speck chip is a system-on-chip combining a dynamic vision sensor and SNN cores containing 320K spiking neurons. It can run large-scale spiking convolutional neural networks (sCNNs) while consuming less than 1~mW of power\cite{yao_spike-based_2024}. 
Dynap-CNN is the processor-only variant of Speck supporting more complex sCNN models with 1 million neurons.
DYNAP-SE2 \cite{richter2024dynapse2}, an analog mixed-signal chip with a built-in biosignal amplifier, is designed for wearable health devices and robotic applications and is more research-oriented.

\subsubsection{BrainChip}
Akida is an event-based processor developed by BrainChip \cite{poseyWhatAkidaEvent2020}. Akida supports a wide variety of neural networks and can execute complex networks. It also supports the AXI bus for connection to CPUs, allowing custom networks not supported by Akida to be executed on the CPU.
It comprises a data processing unit to preprocess input data, converting it into events, and uses an LPDDR4 interface for storing programs and parameters. Additionally, the PCIe interface can be used to connect to other Akida chips.
Akida aims to support a broad range of applications, including robotics and automation in industry, real-time sensing in automotive, vital-signs prediction in on-device health monitoring, and intelligent automation in homes.

\subsubsection{GrAI Matter Labs}
The GrAI VIP chip consists of a CPU and a GrAICore with 196 NeuronFlow cores connected through an event-based network-on-chip (NoC), which is equipped with high-speed interfaces for cameras, microphones, speakers, and the host system.
It is also optimized for both recurrent and feedforward models by supporting 16-bit floating-point data format and is appropriate for edge AI applications, such as audio and video processing \cite{moreiraNeuronFlowNeuromorphicProcessor2020}.

\subsubsection{Intel Loihi 1 \& 2}
Intel Loihi 1 is a programmable digital many-core neuromorphic system that approximates the behavior of biological neurons \cite{daviesLoihiNeuromorphicManycore2018}.
Loihi 2, the 2nd generation chip, comprises 128 neuron cores, each containing 8,192 neurons and 192 kB of memory that can be flexibly allocated between neurons and synapses. Therefore, each chip includes 1 million fully programmable neurons and 120 million synaptic connections. The neuron cores are interconnected by a NoC and support spike-based communications. The chip includes an inter-chip communication interface to facilitate the creation of large 3D chip clusters. %
\cite{daviesTakingNeuromorphicComputing2021}.
Due to the promising scale-up ability, Loihi chips exhibit a big potential for integration into data centers.

\subsubsection{Tianjic}
The Tianjic chip is based on a 156-core architecture with localized memory and streamlined dataflow, which can be used to simulate 40,000 neurons and 10 million synapses \cite{pei2019towards}. The chip, %
supports both artificial neurons and spiking neurons, enabling emulation of various neural networks such as MLP, CNN, and RNN \cite{Pei2023MultigrainedSI}. In contrast to other hybrid chips, such as SpiNNaker2, which have the flexibility to build state machines using non-neural code, the Tianjic chip uses a neural state machine to assemble its applications, trading off flexibility by high integration.

\subsubsection{BrainScaleS-1 \& 2}

BrainScaleS-1 and BrainScaleS-2 (BSS) are analog mixed-signal neuromorphic chips developed by the University of Heidelberg. 
BSS-1 is not a programmable chip, while BSS-2 contains programmable synaptic connections.
BSS-2 contains four analog neuron cores and digital synaptic arrays connected with a spike router. Each analog core includes 512 spiking neurons and 32,768 synapses. A BSS-2 system comprises multiple single-chip setups, which are interconnected to a computing cluster via Ethernet and suitable for robotic applications \cite{schmittNeuromorphicHardwareLoop2017,MULLER2022790,EXTOLLreferenceBrainScale,muller2022scalableSWBrainscale}.

\subsubsection{IBM TrueNorth and NorthPole}
As the pioneer of the brain-inspired AI chips, 
TrueNorth \cite{merolla2014million} incorporates programmable digital neurons, whereas NorthPole \cite{modhaNeuralInferenceFrontier2023} comprises computation units to simulate biological neurons.
NorthPole contains 256 cores interconnected by two dense NoCs.
Inspired by the brain's structure, one is designed for short-distance communication between nearby cores, and the other facilitates long-distance neuron activation communication across all cores.
It contains 1 million programmable neurons and 256 million programmable synaptic connections. In total, 224 MB of on-chip memory is distributed across the 256 cores. The vector-matrix multiplier (VMM) can execute computations in 8-bit, 4-bit, and 2-bit fixed-point data formats. %
It is suitable for image classification, detection, segmentation, natural language processing, and speech recognition \cite{merolla2014million,modhaNeuralInferenceFrontier2023}. 

\subsubsection{SpiNNaker 1 \& 2}
SpiNNaker1 is a custom ARM-based 18-core chip developed in 130nm process technology by the University of Manchester, featuring a massively parallel architecture designed for large-scale real-time brain simulations with spiking neural networks. It %
currently holds the record for the world's largest neuromorphic supercomputer, including a total of 1'036.800 million cores, arranged in 1,200 48-node boards highly interconnected in a toroidal mesh. Such a supercomputer has the potential to emulate roughly 1 billion neurons and 1000 billion synapses, which might vary depending on the neuron models used \cite{spinnworld,SpiNNaker2020book}. 

The successor SpiNNaker2 chip was developed in 22nm FDSOI by TU Dresden and the University of Manchester within the Human Brain Project. SpiNNaker2 features 152 ARM-based processing elements (PEs) for flexible software-based execution of neural networks. SpiNNaker2 deploys the same event packet routing as SpiNNaker1 and is designed to be scaled up to 10 million cores \cite{Mayr2019}.
In addition to scalable brain simulations, SpiNNaker2 also targets efficient real-time AI processing with event-based DNN and generic computation \cite{SpiNNaker2EventBasedML2024}.
SpiNNaker2 has custom accelerators to speed up the processing of DNN layers and for compute-heavy operations in neuromorphic computing \cite{Hoeppner2022}. %

\subsection{Summary of neuromorphic systems}
Multifarious neuromorphic systems cover a wide range of AI applications, from ultra-low power tinyML tasks on the edge to large-scale brain simulations on the cloud.
In this paper, we focus on the neuromorphic systems that can be loaded into data centers for cloud computing.
Those cloud neuromorphic systems are dominated by digital technology and tend to use a more advanced process node for higher power efficiency. Only Innatera, SynSense Dynap-SE2, and BrainScaleS adopt mixed-signal solutions that are restrained in large-scale distributed AI applications due to varied analog process errors.
Tuning AI models is regarded as one of the significant tasks in data centers. However, some cloud neuromorphic platforms such as Tianjic, Truenorth, and NorthPole don't provide the training functionality and focus on inference only.
Moreover, the current state of software frameworks for neuromorphic computing exhibits fragmentation, with each player developing its own software stack to be maximally adapted to its hardware. %

\section{Comparing neuromorphic to conventional solutions for AI processing}
\label{sec:comparing}

\begin{table*}
    \centering
    \caption{Performance of neuromorphic solutions on diverse applications.
    The \emph{energy ratio} is the energy of the conventional divided by the energy of the neuromorphic solution. Analogously, the \emph{time ratio} describes how much faster the neuromorphic solution is. MSE: mean squared error. PPL: perplexity. $\uparrow:$ higher value is better. $\downarrow:$ lower value is better. Tianjic results for VGG16 on ImageNet.}
    \renewcommand{\arraystretch}{1.3}
    \label{tab:comparing}
    \begin{tabular}{ll>{\centering\arraybackslash}m{4cm}m{3.8cm}m{8mm}m{7mm}ll}
        \toprule
        \textbf{Reference} & \textbf{Hardware} & \textbf{Application} & \textbf{Task performance} \par neurom. $\mid$ convent. & \textbf{Energy ratio} & \textbf{Time ratio} & \textbf{Ref. hardware} \\
        \hline
        Deng~2020\cite{deng2020tianjic} & Tianjic & Image Classification (CNN) & $70.83\% |$ n.a. (Top-1 Acc. $\uparrow$) & 39 & 76 & NVIDIA V100\\
        Ceolini~2020\cite{ceolini2020hand} & Loihi & EMG+DVS gesture recognition (CNN) & 96.0\% $\mid$ 95.4\% (Top-1 Acc. $\uparrow$) & 29.1 & 0.89 & Jetson Nano \\
        Patel~2021\cite{patel2021spiking} & Loihi & Image segmentation (U-Net) & $92.13 \% \mid 94.98 \%$ (Pixel Acc. $\uparrow$) & 3.00 & 0.011 & GeForce RTX 2080 \\
        Rao~2022\cite{rao2022long} & Loihi & Time-series classification (RelNET) & 16/17 $\mid$ 16/17 (tasks solved $\uparrow$)& 4.36 & 0.73 & GeForce RTX 2070 \\
        Shrestha~2023\cite{shrestha2023efficient} & Loihi 2 & Video processing (PilotNet) & 0.035 $\mid$ 0.025 (MSE $\downarrow$) & 150 & 2.8 & Jetson Orin Nano \\
        Nazeer~2023\cite{nazeer2023language} & SpiNNaker2 & Language modelling (EGRU) & 97.3 $\mid$ 97.3 (Test PPL $\downarrow$) & 18.3 & 0.117 & NVIDIA A100 \\
        \bottomrule
    \end{tabular}
\end{table*}

\begin{figure}[htbp]
    \centering
    \includegraphics[width=1.0\linewidth]{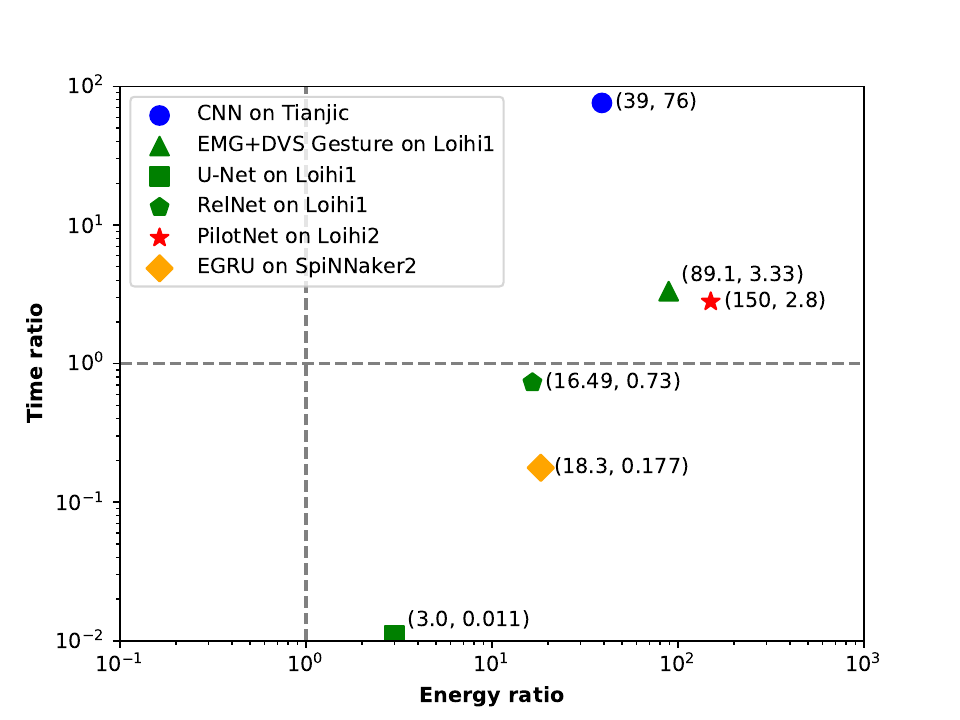}
    \caption{Comparison of energy and solution time ratios from \cref{tab:comparing}
    }
    \label{fig:comparing}
\end{figure}

\subsection{Comparison of energy and speed}
Targeting the integration of neuromorphic hardware into data centers for efficient AI processing, we next want to evaluate \emph{for which applications} there are neuromorphic solutions and \emph{by how much} these solutions are better in terms of energy and latency compared to conventional solutions.
To do so, we have reviewed articles where the hardware systems from \cref{sec:hardware} are benchmarked against conventional solutions (CPUs, GPUs, or accelerators like TPUs) in terms of energy, latency, and accuracy on the AI task.
We neglect results on rather simple tasks like handwritten digit recognition or keyword spotting and instead look at more complex tasks as our focus is on AI deployment in data centers.

Six exemplary results on four hardware platforms are shown in \cref{tab:comparing} and \cref{fig:comparing}: 
For each example, we report how much more energy-efficient the neuromorphic solution is (energy ratio) and how much faster it is to the compared solution (time ratio). In all cases, the correctness of the solution is comparable on both neuromorphic and conventional hardware.
Regarding energy per inference, Loihi 1 defeats GPUs for relational reasoning with spiking LSTM \cite{rao2022long}, image segmentation \cite{patel2021spiking}, and multi-sensor spatio-temporal gesture classification \cite{ceolini2020hand} by factor of 3 to 100. Loihi 2 achieves more than $100\times$ better energy on the automotive PilotNet benchmark with sigma-delta networks \cite{shrestha2023efficient}. Tianjic requires on average $39\times$ less energy than a GPU for image classification using non-spiking CNNs \cite{deng2020tianjic}. SpiNNaker2 achieves an $18\times$ energy improvement for language modeling with event-based GRU \cite{nazeer2023language} versus a data center GPU.

Concerning speed, there is no clear winner between neuromorphic systems and GPUs. Depending on the task and system, the neuromorphic solution can be up to $100\times$ faster or $100\times$ slower than on GPUs. Note that all time results in \cref{tab:comparing} and \cref{fig:comparing} are given for a batch size of 1. For larger batch sizes, GPUs achieve a lower average time per inference at improved energy efficiency compared to batch size 1, while on neuromorphic systems, the average time and energy per inference remain constant  \cite{patel2021spiking,shrestha2023efficient,nazeer2023language,rao2022long}. Hence, at larger batch sizes, the advantage of neuromorphic systems decreases.

\subsection{Interim conclusions}
From the above results, we draw some interim conclusions:
\begin{itemize}
    \item Tasks and models: Neuromorphic solutions are available for image processing with CNN, natural language processing with recurrent neural networks (spiking LSTM or event-based GRU), and spatiotemporal pattern recognition.
    \item Energy: Neuromorphic hardware is between 3 to 100 times more energy efficient per inference at batch size~1.
    \item Speed: For some tasks, neuromorphic hardware shows faster inference compared to conventional systems. This advantage diminishes for larger batch sizes.
\end{itemize}

\subsection{Limitations of analysis}
In our analysis, we consider the energy ratio and solution time ratio compared to benchmarked GPU in the corresponding publications. 
It is clear that new conventional AI hardware platforms are available today that likely show a lower energy and lower inference time. Accordingly, an alternate approach would be to compare neuromorphic solutions to current state-of-the-art benchmark results.
However, it would be unfair to benchmark a 5-year-old neuromorphic chip to a brand-new edge GPU using the latest software technology. 
Hence, we here defend our approach as both neuromorphic and conventional solutions represented the state-of-the-art at the time of publication.

We further note that the list of presented neuromorphic solutions is very limited and non-exhaustive. Due to a lack of public benchmarking results fulfilling our criteria, we cannot highlight any results from IBM TrueNorth or commercial neuromorphic system providers (Innatera, SynSense, BrainChip, and GrAI Matter Labs).  

For a better and more transparent comparison between neuromorphic and conventional AI platforms, we recommend submitting results for standard machine learning benchmarks to MLPerf \cite{reddi2020mlperf} and to the newly established neuromorphic benchmark NeuroBench \cite{yik2024neurobench}.

\section{Applications, models, and algorithms}
\label{sec:applications}
This section explores relevant AI workloads in diverse industrial applications, followed by an examination of the current state of spiking neural networks and their applications in machine learning. Finally, we outline future research directions in this evolving field.
\subsection{AI Workloads in Industries and Research}
\subsubsection{Production Industry Applications}
The industry will be under enormous pressure to transform itself to a carbon-neutral future in the coming years. A good example of this is the steel industry, which wants to significantly reduce its $CO_2$ emissions as compared to 1990 ~\cite{lcr}. To meet these challenges, technologies, and tools must be created at TRL8 (Technical Readiness Level)~\cite{estepclean}. Of course, it is important to ensure that methods based on new technologies are also energy-efficient to avoid rebound effects. 

From an industrial point of view, typical applications for AI models are computer vision-related tasks. For tasks like defect detection, tracking, ID recognition, or anomaly detection, camera-based solutions are fairly common. These types of tasks tend to be operated 24/7 with high throughput (production lines are often operated at speeds $>$ 1m/s) and high workload, as defects can be fairly small and fairly local, making rigorous requirements on the high-resolution input images or video streams. So these real-time inputs are often evaluated 24/7 on several GPUs by multiple AI models at various production stations, making the energy consumption of the productive system non-negligible.

\subsubsection{Digital Industry Applications}

AI has shown super-linear growth trends in its share of computing usage in data centers \cite{andrae2015global}. The most common applications for AI in the digital industry include recommendation models, language models, vision models, etc., and it will most likely become even more prevalent in the future. Moreover, in recent years, generative models such as Large Language Models (LLM) and Large Multimodal Models (LMM) have gained explosive growth \cite{Sevilla2022}. In these applications, a significant portion of computing resources is dedicated to inference, given their role as generative services for end-users. As a result, energy consumption becomes a crucial consideration for the models.

\subsubsection{AI workloads in Research}
Looking at deep learning research, we see that the following AI model types are still relevant: convolutional neural networks, transformers, graph neural networks, generative adversarial networks, variational autoencoders, normalizing flows, diffusion models and deep reinforcement learning \cite{prince2023understanding,bishop2023deep}.
From the models above, neuromorphic computing has mainly looked into convolutional networks (see \cref{sec:comparing}) and into reinforcement learning \cite{wunderlich2019demonstrating}.
Recurrent neural networks, for which SNN and neuromorphic computing have shown very efficient solutions \cite{daviesTakingNeuromorphicComputing2021}, have moved out of focus a bit.

\subsubsection{Need for AI workload statistics}
From the academic literature and from general media it is possible to extract which AI applications and which AI models are trending. Also, it is now common to report the cost for training AI models in the machine learning literature as the GPU hours used \cite{neurips2024ethics}, or even to provide the estimated $CO_{2}e$ emissions \cite{anthony2020carbontracker}. Unfortunately, we miss the public information about which AI models are run, how often in commercial data centers, and what is their share of the total compute resources. The big tech companies only share rough information, e.g., about the relative increase of AI tasks or the share of inference, training, and network architecture search \cite{wu2022sustainable}. 
Detailed AI workload statistics would help determine which AI tasks and models to focus on for developing energy-efficient neuromorphic solutions. Finally, this could help to apply the greatest leverage for reducing the operational cost in AI data centers.

\subsection{Spiking Neural Networks in Machine Learning: Current Landscape}
Spiking neural networks (SNN) \cite{maass1997networks} are well-suited for efficient event-based implementations and have been scaled to very large sizes \cite{jordan2018extremely}. However, for many years, they have been avoided in machine learning because of their non-differentiable dynamics, which at first glance made them ill-suited for gradient-based learning. However, several methods have been proposed to tackle this problem, enabling gradient-based end-to-end learning for SNNs \cite{bellecSolutionLearningDilemma2020, Wunderlich2021, chen2018neural}. This development has led to several successful implementations of SNN learning on neuromorphic hardware. Algorithms like Spiking-based Backpropagation \cite{neftci2019surrogate}\cite{zhu2022training} and other hybrid methods have been shown to achieve comparable performance as their ANNs counterparts at image classification tasks \cite{fang2021incorporating}\cite{fang2021deep}. Additionally, training techniques like Spiking Generative Networks in Lifelong Learning Environment \cite{zhang2023spiking} demonstrated their effectiveness in image classification and generation.

The most powerful SNN models that reach close to the state-of-the-art performance of conventional machine learning models have so far only been demonstrated in software. Recently, there has been increasing interest in porting large transformer-based models \cite{vaswani2017attention} to SNNs \cite{zhou2022spikformer, zhu2023spikegpt, lv2023spikebert, bal2023spikingbert, zhou2024spikformer}. Several recent studies have highlighted the high levels of sparsity in transformers \cite{li2022event, liu2023deja, jaszczur2021sparse, han2023hyperattention, li2023lazy, correia2019adaptively, fedus2022switch}, and spiking neurons are an efficient method to make use of this feature. Specifically, Spikformer \cite{zhou2022spikformer} is a variant of the transformer network architecture based on spiking neural networks. They introduce a spiking variant of self-attention to efficiently implement transformers with SNNs and demonstrate compelling performance on a range of benchmark tasks. Recently, a newer version of this model has been published, which reaches up to the performance of its non-spiking counterparts at a significantly reduced compute budget \cite{zhou2024spikformer}. In addition, other research groups have introduced spiking variants of specific popular transformer architectures for large language models, such as GPT \cite{zhu2023spikegpt} or BERT \cite{lv2023spikebert, bal2023spikingbert}, demonstrating promising results.

\subsection{Research Directions for SNNs in Machine Learning}
As we approach the forefront of the current research, we suggest that the future lies in incorporating SNN-based models into mainstream applications, focusing on extending large-scale applications beyond image classification, refining training techniques, and optimizing mapping strategies for diverse hardware platforms. Efforts in standardization and compatibility with existing frameworks will enhance SNN adoption across industries. Additionally, delving into novel applications, particularly in real-time scenarios, and further integration with transformer-based models could unlock new frontiers for SNN research. Emphasizing energy efficiency, economic viability, and scalability will be pivotal for solidifying SNNs as a key player in the evolving landscape of machine learning and neuromorphic computing.

\section{Hardware and software integration}
In this section, we discuss the hardware and software challenges for integrating neuromorphic hardware platforms into AI data centers.

\subsection{Hardware integration}
\label{sec:hardware}

\subsubsection{Status quo}
So far, the following neuromorphic systems have been integrated at large scale into data centers: SpiNNaker~1 \cite{furber2016large}, TrueNorth (NS16-4e) \cite{debole2019truenorth}, Loihi (Pohoiki Springs) \cite{frady2020neuromorphic}, BrainScaleS-1 \cite{schmidt2023clean} and Tianjic \cite{Pei2023MultigrainedSI}.
All systems use slide-in modules with custom printed circuit boards (PCBs) for integration into standard 19" server racks. Typically, the neuromorphic chips are accessed via Ethernet, only the TrueNorth NS16e-4 uses PCIe for communication with the host chip. Baseboard Management Controller (BMC) or similar controllers are used for booting and monitoring the boards. All platforms also include field-programmable gate arrays (FPGAs) or system-on-chips (SoCs), most often as middleware between host computers and neuromorphic systems.
Some systems have already integrated a host CPU, e.g., Pohoiki Springs or the Tianjic server, while the other systems require external host CPU servers for the configuration and control of the neuromorphic systems and for preprocessing.

As an exception, BrainChip offers PCIe boards for integrating their Akida chips with CPU servers. This represents another option for integrating neuromorphic computing systems into data centers, similar to normal GPUs. Note, however, that this might limit the size of neuromorphic models that can be implemented compared to the larger systems discussed above.

\subsubsection{Conclusion}
The above examples show that a variety of neuromorphic systems have been successfully integrated into standard data center server racks. Thus, technically, the hardware integration does not pose a problem. Yet, we observe a diversity in how large neuromorphic systems are assembled into server boards, e.g., many of them leverage FPGAs or SoCs as middleware. These extra devices and the host CPU add a power overhead to the very energy-efficient neuromorphic systems. Optimizing for system-level efficiency of AI compute servers, these components need to be included when performing benchmarking on AI workloads.
Another requirement for the industry-level deployment of neuromorphic chips is high reliability and robustness. The chips and boards need to be designed for a 24/7 operation, e.g., the server board should keep working if a single chip or processor fails. Replacement parts should be available for a long period.

\subsection{Software and operation}
\label{sec:software}
\subsubsection{Operation principles}
The way how neuromorphic systems are operated significantly differs from other AI accelerators or GPUs. There is no operating system or runtime that schedules compute tasks sequentially on neuromorphic cores. Instead, the synaptic weights and neuron parameters are configured first on all cores and chips of the neuromorphic systems. Then, input data, such as spikes or scalar events, are streamed into the NC system and processed by the neurons and synapses. Most of the systems operate in real-time, which means the individual chips run asynchronously, and spike events are processed as they arrive. In mixed-signal systems, the decay time constants of the silicon neurons define the speed of operation, which may be equal to the speed of biological neurons like in DYNAP-SE2\cite{richter2024dynapse2}, or accelerated by a factor of 1000 to 10000 in the BrainScaleS systems \cite{muller2022scalableSWBrainscale}. Digital systems such as TrueNorth, SpiNNaker, or Loihi typically split the neuron updates into timesteps. How long it takes to process one timestep then defines how fast a spiking network can be executed. Loihi offers a barrier synchronization to continue with the next time step once all cores in all jobs are done with the current step. In contrast, in SpiNNaker, neuron updates are triggered in regular intervals (e.g., 1~ms), effectively yielding a real-time system. Typically, for AI inference neuromorphic systems process with batch size 1. Multiple inputs are processed sequentially or need to be distributed spatially onto different system cores. Note, however, that some SNN architectures allow for pipelining input data when each layer requires the same fixed number of timesteps \cite{esser2016convolutional,lopez2022conversion}.

The real-time operation of neuromorphic systems poses a challenge for integration into cloud-based digital computing systems. Nilsson et al.~\cite{nilsson2023integration} frame this challenge in detail and suggest a conceptual framework based on micro-services to solve it.
For data centers, we consider several scenarios on how AI compute load arrives at the host CPU managing the neuromorphic system:
\begin{itemize}
    \item Streaming application in edge cloud, e.g., for processing data from a video camera at 30 FPS in real-time. In this case, a neuromorphic system is pre-configured with a neural network model. Each image needs to be converted to data suitable for the hardware (e.g., spikes) and then provided into a neuromorphic system. Results are recorded and processed further on the host CPU. Depending on the SNN model, the states need to be reset after each input image. Another option might be to use a network model for video streaming that does not require an external state reset.
    \item Irregular requests from the web. The frequency of AI compute requests arriving in the data center may vary across the time of the day, the week, or even seasons. This creates very irregular workload statistics. Data centers with neuromorphic hardware will need to handle such scenarios. Again, for a specific task, the neural network weights can be pre-loaded. Then, inference is triggered by arriving requests. For neural network architectures supporting pipelining, multiple requests can be buffered and forwarded sequentially into the NC hardware to achieve the highest possible throughput. Maximum response latencies need to be adhered to. If the incoming request rate becomes too high, the processing of the AI tasks needs to be redirected to other neuromorphic systems or conventional compute units. In longer idle intervals, the neuromorphic chips may even be switched off or sent into sleep or retention mode.
    \item Regular, low-priority AI tasks that can be scheduled to free neuromorphic resources. For such tasks, the execution typically also includes the generation and loading of the hardware configuration. There is no specific challenge except for achieving a high utilization of the neuromorphic chip resources.
\end{itemize}
We note that for most of the tasks, a pre-processing of data (e.g., encoding into spikes) and the post-processing of data (e.g., decoding of spike rates into class probabilities) are required

Regarding the dynamic scheduling of AI compute jobs onto neuromorphic hardware, data center standards need to be adopted. Current cloud-scale neuromorphic systems either use SLURM \cite{mueller2022operating} or custom schedulers \cite{rowley2019spinntools}, while commercial data centers use container orchestration platforms like Kubernetes or alternatives.

\subsubsection{High-level software}
To enable neuromorphic computing for efficient AI processing, advanced software tools are needed for the training and deployment of SNN or other brain-inspired models \cite{qu2022review}. While there exists a multitude of software frameworks for training SNN in PyTorch \cite{eshraghian2021,norse2021,fang2023spikingjelly} or Jax \cite{kade_heckel_spyx}, deep SNN needs to be further optimized for each hardware platform considering details of the neuron model or quantization of weights. Because of this, each neuromorphic system provides its own software framework, as is shown in the last column of \Cref{tab:systems}. While there are some approaches for unifying the programming model of NC \cite{Aimone_2019,Lava_2023,pedersen2023neuromorphic}, none of them have evolved to a standard yet.

However, the standardization of tools and ease of use are key factors for a new technology to obtain acceptance in the industry. Thus, to proliferate SNNs further in an industrial setting, standard APIs for developing and training SNNs are required, e.g. comparable to Keras~\cite{keras}, TensorFlow~\cite{tf} or PyTorch~\cite{pytorch}, ideally even a frontend such as Keras with the backend, whether it be SNN-based or a conventional TensorFlow or PyTorch, freely interchangeable. This would make development on standard hardware possible and an easy transition to specialized SNNs for a productive system, regardless of whether the SNNs are located on-premise or in an external data center, aka in the cloud. This kind of portability could also help avoid the fear of vendor lock-in when transitioning to a comparably new technology stack, as well as questions regarding long-term support when switching from established suppliers to a new, comparably small hardware supplier.

In addition to that, it is essential to provide a large number of examples and a so-called model zoo of validated AI models for each hardware platform. So far, only BrainChip provides a model zoo for their Akida systems \cite{brainchip2024modelzoo}. This allows us to retrain or fine-tune existing model architectures to customers' needs, which not only offers a shorter time-to-solution but also reduces the environmental footprint, as training from scratch can be avoided.

\section{Conclusion}
\label{sec:outlook}
In this article, we reviewed neuromorphic hardware platforms and algorithms for their suitability in reducing energy consumption in AI data centers. We also discussed the current challenges that neuromorphic computing faces in becoming a mainstream technology used by the industry. In particular, we analyzed that the current AI model types supported by neuromorphic computing only partially match the AI models commonly run in AI data centers. We conclude that the neuromorphic computing community should focus on state-of-the-art ML technologies, such as transformers, and needs to establish standardized software frameworks that ensure interoperability among hardware vendors.

Data center sustainability is not only about saving energy during operations but also about saving water and materials while keeping social and governance issues in mind. These latter issues are becoming increasingly important as AI models use vast amounts of personal data. %
When considering the carbon footprint, one will eventually face the question of whether or not to integrate specialized hardware: the embodied footprint of an additional device may be greater than the operational footprint savings due to specialized solutions \cite{eeckhout2022first}. Neuromorphic engineers should therefore focus on the high utilization of their platforms.

\bibliographystyle{IEEEtran}
\bibliography{references}

\end{document}